\documentclass[journal=jacsat,manuscript=article]{achemso}

\usepackage{chemformula}
\usepackage[T1]{fontenc}
\usepackage{comment}
\usepackage{algorithm}
\usepackage{rotating}
\usepackage[noend]{algpseudocode}
\usepackage{tikz}
\usetikzlibrary{shapes, arrows}
\usepackage{caption}
\usepackage{subcaption}
\usepackage{subcaption}
\usepackage[font=small]{caption}
\usepackage[labelformat=empty, position=top]{subcaption}
\usepackage{cancel}

\usepackage{amsfonts}
\usepackage{ulem}

\author{Siyavash Moradi}
\affiliation{Technical University of Munich; TUM School of Natural Sciences and Catalysis Research Center, Department of Chemistry, Lichtenbergstr. 4, 85748 Garching, Germany}
\altaffiliation{These authors contributed equally.}

\author{Rebecca Tomann}
\affiliation{Pitzer Center for Theoretical Chemistry, Department of Chemistry, University of California, Berkeley CA 94720, USA}
\altaffiliation{These authors contributed equally.}

\author{Josie Hendrix}
\affiliation{Pitzer Center for Theoretical Chemistry, Department of Chemistry, University of California, Berkeley CA 94720, USA}

\author{Martin Head-Gordon}
\affiliation{Pitzer Center for Theoretical Chemistry, Department of Chemistry, University of California, Berkeley CA 94720, USA}

\author{Christopher J. Stein}
\affiliation{Technical University of Munich; TUM School of Natural Sciences and Catalysis Research Center, Department of Chemistry, Lichtenbergstr. 4, 85748 Garching, Germany}

\email{christopher.stein@tum.de, m_headgordon@berkeley.edu}

\title
  {Spin parameter optimization for spin-polarized extended tight-binding methods}

\begin{document}

\begin{abstract}
\noindent We present an optimization strategy for atom-specific spin-polarization constants within the spin-polarized GFN2-xTB framework, aiming to enhance the accuracy of molecular simulations. We compare a sequential and global optimization of spin parameters for hydrogen, carbon, nitrogen, oxygen, and fluorine. Sensitivity analysis using Sobol indices guides the identification of the most influential parameters for a given reference dataset, allowing for a nuanced understanding of their impact on diverse molecular properties. In the case of the W4-11 dataset, substantial error reduction was achieved, demonstrating the potential of the optimization.
Transferability of the optimized spin-polarization constants over different properties, however, is limited, as we demonstrate by applying the optimized parameters on a set of singlet-triplet gaps in carbenes.   
Further studies on ionization potentials and electron affinities highlight some inherent limitations of current extended tight-binding methods that can not be resolved by simple parameter optimization. We conclude that the significantly improved accuracy strongly encourages the present re-optimization of the spin-polarization constants, whereas the limited transferability motivates a property-specific optimization strategy.
\end{abstract}

%----------------------------------------------------------------
%Motivation
\section{1. Motivation}\label{Introduction}

Semi-empirical electronic-structure methods offer an attractive balance between computational efficiency and accuracy. This enables the study of large and complex systems that are computationally too demanding for more accurate methods, such as density functional theory (DFT)\cite{argaman2000density,chermette1998density, kohn1996density}. The density functional tight-binding method (DFTB)\cite{elstner1998self} is such an approach that significantly reduces the computational cost of DFT, making calculations using standard hardware on large systems\cite{elstner2000self,frauenheim2000self} with several hundred to a few thousand atoms feasible. This method allows for exploration on longer time scales, enabling the investigation of dynamical properties, as recently exemplified by a simulation of the characteristic barriers of a $\mathrm{Au^{-}_{14}}$ -- cluster at high-temperature 
\cite{koskinen2007liquid}. Moreover, DFTB is valuable for structural
exploration\cite{jackson2004unraveling} and as a preliminary screening tool for subsequent DFT calculations\cite{koskinen2006density,koskinen2008self}. DFTB compares favorably to full DFT with small basis sets. It is particularly well-suited for covalent systems like hydrocarbons\cite{porezag1995construction} and performs surprisingly well in describing metallic bonding with delocalized valence electrons\cite{kohler2005density}.
%\subsection{recent success of fast electronic structure methods}\label{apb}
The success of this relatively fast electronic-structure method has prompted further in-depth studies in a variety of target research areas, such as the exploration of chemical reactions and reaction kinetics\cite{he2019reaction, hamilton2021predicted}. Additionally, these methods have been instrumental in investigating materials properties such as band structures\cite{kurban2020tailoring} and optical properties\cite{alkan2018td}, enabling the design of novel materials with tailored functionalities.

%\subsection{extended tight binding models are particulary successful}\label{apb}
%(what gfn2xTB does than dftb)
\noindent Among the various fast electronic-structure methods, extended tight-binding (xTB) models have gained particular prominence\cite{bannwarth2021extended}. These models approximate the electronic structure using a minimal basis set, capturing the essential quantum-mechanical interactions at low computational cost. 
By incorporating elements of valence bond and molecular orbital theories, xTB models can describe a wide range of chemical bonding types and accurately compute ground-state properties such as bond lengths, angles, and vibrational frequencies. GFN2-xTB\cite{bannwarth2019gfn2}, an extension of the original GFN-xTB\cite{grimme2017robust} model, provides improved accuracy in reproducing experimental data and benchmark results, owing to advancements in the parameterization,  the inclusion of anisotropic electrostatics, and the density-dependent D4 dispersion correction\cite{caldeweyher2019generally}. It mostly retains the computational efficiency of its predecessor.
The spin-polarized DFTB method (SDFTB)\cite{frauenheim2000self} has been successfully applied to investigate e.g. metallic clusters, such as the Fe$_{147}$ icosahedron\cite{kohler2007treatment}. 
The recently published spin-polarized GFN2-xTB (spGFN2-xTB) method \cite{neugebauer2023high} was benchmarked on main-group thermochemistry datasets including open-shell molecules.
The spin-polarization parameters are obtained from PBE calculations (vide infra) and typically not re-optimized in the tight-binding framework. Adopting these parameters directly from another theory is unusual since the tight-binding parameters are typically obtained from fitting to reference data.
In this article, we will explore potential gains in accuracy upon re-optimization of the spin-polarization parameters. We will further develop a strategy that allows us to determine which parameters can be optimized with a given reference dataset. This procedure relies on a sensitivity analysis by means of Sobol indices \cite{sobol1993sensitivity}. We then analyze the transferability of the re-optimized parameters and discuss optimization strategies for different application scenarios.

\section{2. Theory}\label{detailes1}
\subsection{2.1 Spin-polarization in extended tight-binding methods}
The total spGFN2-xTB energy expression which has been recently implemented in the Q-Chem software package\cite{krylov2013q,epifanovsky2021software} (QChem-xTB) is given by

\begin{equation}\label{eq:total}
\begin{aligned}
E_{\mathrm{QChem-xTB} }= & E_{\text {rep }}+E_{\mathrm{EHT}}+E_{\mathrm{IES}+\mathrm{IXC}}+E_{\mathrm{AES}} \\
& +E_{\mathrm{AXC}}+E_{\mathrm{spin}}
\end{aligned}
\end{equation}

\noindent and is similar to the one in Ref.~[22] with the exception of the Fermi and D4 terms. The abbreviations refer to the repulsion (rep), extended Hückel-type (EHT), isotropic electrostatic (IES) and isotropic exchange-correlation (IXC), anisotropic electrostatic (AES) and anisotropic exchange-correlation (AXC) energies and the spin-polarization contribution.\\
We consider the dependence of the energy on the spin densities with a spin-polarization term,\cite{frauenheim2000self,kohler2005density,kohler2001approximate} which has the following form
\begin{equation}
E_{\mathrm{spin}}\approx \frac{1}{2} \sum_A^N \sum_{l \in A} \sum_{l^{\prime} \in A} p_{A l} p_{A l^{\prime}} W_{A l l^{\prime}}
\label{eq: spin_term}
\end{equation}

\noindent where $p_{Al}$ represent the differences between $\alpha$ and $\beta$ Mulliken populations for an orbital on atom $A$ with angular momentum $l$. \\
\noindent The universal atomic constants  $W_{All^{\prime}}$ --- the spin-polarization parameters --- can be computed for free atoms by determining the second derivatives of the DFT total energy expressions in relation to the magnetization density. Using  Janak's theorem\cite{janak1978proof}, we can express these parameters via the dependence of the orbital energies on the orbital occupation.
%These derivatives correspond to differences between occupation numbers(Using Janak's theorem\cite{janak1978proof}).

\begin{equation}
W_{All^{\prime}}=\frac{1}{2}\left(\frac{\partial \varepsilon_{l \uparrow}}{\partial n_{l^{\prime} \uparrow}}-\frac{\partial \varepsilon_{l \uparrow}}{\partial n_{l^{\prime} \downarrow}}\right)
\label{eq: Ws}
\end{equation}

\noindent where the $n_{l^{\prime} \uparrow}$ and $n_{l^{\prime} \downarrow}$ are the atomic occupation numbers for alpha and beta electrons, respectively, and the $\varepsilon_{l \uparrow}$ are the Kohn--Sham eigenvalues.

\noindent The Hamiltonian matrix elements for the spin-polarization term are given by 

\begin{equation}
H_{\mu \nu} = \pm \frac{1}{2} S_{\mu v} \sum_{l^{\prime \prime} \in A}\left(W_{A l^{\prime \prime}}+W_{A l^{\prime} l^{\prime \prime}}\right) p_{A l^{\prime \prime}}
\end{equation}

\noindent where $S_{\mu \nu}$ is the overlap matrix and the plus and minus signs correspond to $\alpha$ and $\beta$ electrons, respectively.\\
%\section{2. Why sensitivity analysis and  What is sobol indices?}\label{Introduction}
\subsection{2.2 Sensitivity Analysis}
Sensitivity analysis quantifies the dependence of an output on variations of the input variables.
Rather than deriving expected outcomes or probability distributions, it provides a range of possible output values linked to each input set.\\
Sobol's technique relies on decomposing the variance of the model output into individual components, each representing the variance of the input parameters at progressively higher dimensions \cite{saltelli1999quantitative, sobol1993sensitivity}. The contribution of each individual input parameter and groups of them to the total variance of the model output is determined. 
It is important to recognize that Sobol sensitivity analysis is not designed to pinpoint the sources of input variability. Instead, it merely reveals the impact and magnitude of such variability on the model output. Consequently, it is unsuitable for identifying the origin(s) of variance. \\ 
The output of the model, for which the sensitivity to input parameters is being evaluated, is $g(\bold{x})$. In the probabilistic interpretation of the parameters, $g(\bold{x})$ is considered a random variable with a mean ($g_{0}$) and variance (V)\cite{zhang2015sobol}

\begin{align} 
 g_0=\int g(\bold{x}) \mathrm{d} \bold{x} \; \; \; \ , \; \; \; \; V=\int g(\bold{x})^2 \mathrm{d} \bold{x}-g_0{ }^2 .
\end{align}

\noindent The Sobol method entails decomposing the variance (V) by assigning contributions to the impacts of individual parameters and the collective impacts of pairs of parameters. This is initially accomplished by decomposing the function $g(\bold{x})$ into
\begin{equation}\label{eq:decompos}
g(\bold{x})=g_0+\sum_{i=1}^s g_i\left(x_i\right)+\sum_{i=1}^s \sum_{i \neq j}^s g_{i j}\left(x_i, x_j\right)+\ldots g_{1 \ldots s}\left(x_1, x_2, \ldots, x_s\right) .
\end{equation}

\iffalse
The components of the decomposition are formulated in the following manner \cite{sobol2001global} :
$$
\begin{gathered}
g_i\left(x_i\right)=\int g(x) \prod_{k \neq i} d x_k-g_0 \\
g_{i j}\left(x_i, x_j\right)=\int g(x) \prod_{k \neq i, j} d x_k-g_0-g_i\left(x_i\right)-g_j\left(x_j\right) .
\end{gathered}
$$
\fi
\noindent The analysis of variance representation for $g(\bold{x})$ relies on the fulfillment of the condition as shown below
\begin{equation}
    \int g_{i 1}, \ldots, i_{i s}\left(x_{i 1}, \ldots, x_{i s}\right) \mathrm{d} x_k=0 \text { for } k=i_1, \ldots, i_s  .
\end{equation}
Due to this characteristic, squaring the sides of Eq. \ref{eq:decompos} and integrating the results leads to the decomposition expression for variance
\begin{equation}
V=\sum_{i=1}^k V_i+\sum_{i<j} V_{i j}+\sum_{i<j<l} V_{i j l}+\cdots+V_{1,2, \ldots, k} \;.
\end{equation}

\noindent The Sobol sensitivity indices are subsequently defined for each specific group of parameters as
\begin{equation}
S_{i_1 \ldots i_s}=\frac{V_{i_1 \ldots i_s}}{V} .
\end{equation}
$S_i=\frac{V_i}{V}$ is the first-order contribution from $i_{t h}$ input parameter to the output variance and $S_{i j}=\frac{V_{i j}}{V}$ is used to compute the additional second-order contribution from the interaction between $i_{\text {th }}$ and $j_{\text {th }}$ parameters. Total order sensitivity indices are defined as the sum of all the sensitivity indices as $S_{T i}=S_i+S_{i j_{i \neq j}}+\cdots+S_{1 \ldots i . . s}$.\\
The first-order sensitivity indices quantify the fractional contribution of an individual parameter to the output variance. Second-order sensitivity indices are utilized to gauge the fractional contribution of interactions among parameters to the output variance. Total-order sensitivity indices encompass  first-, second-, and higher-order effects, necessitating an evaluation across the entire parameter space. Higher sensitivity index values indicate a greater influence of corresponding model parameters.\\
Performing sensitivity analysis, particularly using methods like Sobol indices, is crucial for understanding the impact of model parameters on the overall performance of computational methods such as spGFN2-xTB. It allows us to identify which parameters significantly influence the model output and, equally important, to what extent. Sobol indices provide a quantitative measure of the contributions of individual parameters and their interactions to the variability in predictions. By conducting sensitivity analysis, we gain valuable insights into the robustness and reliability of a given model, helping to prioritize efforts in refining parameters that play a pivotal role. This knowledge not only enhances the accuracy of predictions but also guides targeted optimizations, ensuring a more efficient and tailored approach to parameter tuning across diverse chemical and physical systems.
%\section{3. detailes of sesitivity analysis?}\label{detailes1}
We use SALib\cite{herman2017salib}, an open-source Python package designed for sensitivity analysis. Because SALib offers a decoupled workflow, it does not interact with the computational or mathematical model directly. Rather, SALib is in charge of creating the model inputs to calculate the sensitivity indices from the model outputs.
%\section{3. detailes of optimization?}\label{detailes1}
\subsection{2.3 Parameter Optimization}
For the optimization of the spin-polarization parameters, we employed the zeroth-order Powell algorithm. Focusing on the most common elements in organic molecules (hydrogen, carbon, nitrogen, oxygen, and fluorine), we initially optimized the parameters to align our computational results with the reference data obtained from the W4-11 dataset\cite{karton2011w4}. In this paper, we will denote the subset of the W4-11 dataset with molecules that only contain the above elements as W4-11-HCONF.\\
Since the spin-polarization term must lead to a decrease in energy, the spin-polarization constants must have negative values, and we chose the bounds [-0.1, 0]~$E_\text{H}$ in our parameter optimization. The lower bound was chosen based on a survey on the parameters published from PBE-derived results of Eq. \ref{eq: Ws} \cite{neugebauer2023high}. This choice also ensures that the total energy is not erroneously dominated by the spin-polarization term.
%\section{4. Details of the implementation in Q-Chem}\label{detailes1}

\noindent \textbf{Implementation in Q-Chem: }In our implementation of a spGFN2-xTB variant, we exploited the framework and libraries of the Q-Chem\cite{krylov2013q,epifanovsky2021software} software. To achieve accurate calculations and efficient convergence during the Self-Consistent Field (SCF) procedure, we apply the DIIS (Direct Inversion in the Iterative Subspace\cite{pulay1980convergence,pulay1982improved}) algorithm. For the definition of reference valence shell occupations, we followed the occupations in the spGFN2-xTB method\cite{bannwarth2019gfn2} with the exception of nitrogen, which we treat with the reference valence shell occupation $2s^22p^3$ instead of fractional reference occupations. Details of the implementation will be described in a forthcoming article. 
\section{3. Results}\label{detailes1}
\subsection{3.1 Atomization and reaction energies}
\begin{figure}[b!]
    \centering
    \includegraphics[width=\textwidth]{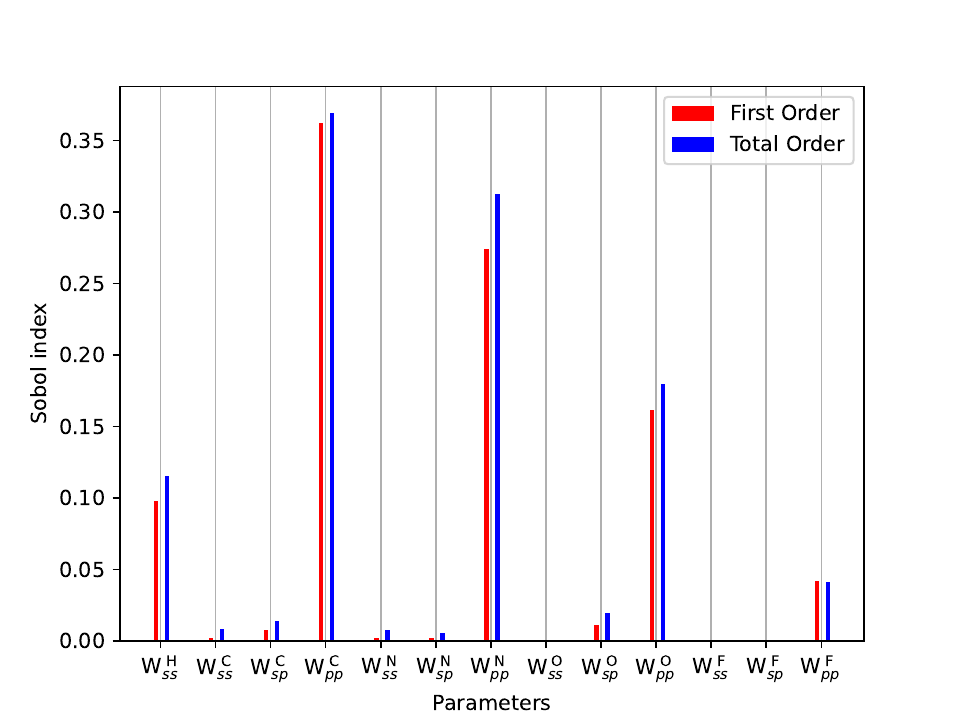}
    \caption{\label{fig:w4_sens} 
    Total-order (blue) and first-order (red) Sobol sensitivity indices of thirteen spin parameters evaluated for the W4-11-HCONF subset.}
\end{figure}

We start with a sensitivity analysis of the spin-polarization parameters for the W4-11-HCONF subset to gain insight into the individual impacts of these parameters for atomization energies. We restrict the optimization to $W_{ss}$ for hydrogen, and $W_{ss}$, $W_{sp}$, and $W_{pp}$ for carbon, oxygen, nitrogen, and fluorine. This initial exploration allows us to unravel the nuanced contributions of each parameter to the overall variability in calculated results. Figure \ref{fig:w4_sens} shows first-order and total-order Sobol indices for the spin-polarization parameters evaluated for W4-11-HCONF.\\
This analysis reveals the importance of the parameters. Notably, $W_{ss}$ for hydrogen and the $W_{pp}$ parameter for the remaining elements have the most influence on the quality of the calculations while $W_{ss}$ and $W_{sp}$ demonstrated minimal impact for these elements. This is to be expected since atomization reactions of closed-shell molecules are studied and the spin-density dependence is determined by the open-shell ground states of the individual atoms.
~
\noindent Table \ref{tab:spin-parameters} shows the results of two optimization strategies. For sequential optimization, spin-polarization parameters were optimized in the sequence hydrogen, carbon, nitrogen, oxygen, and fluorine and the parameters that were already optimized were held constant. This step-wise scheme generally simplifies the optimization problem. 

\begin{table}
  \centering
  \caption{Optimized spin parameters for  W4-11-HCONF subset from two different optimization approaches and default parameters taken from Ref.~\citenum{neugebauer2023high}.}
  \label{tab:spin-parameters}
  \begin{tabular}{|l|c|c|c|}
    \hline
      & Default $W$& Sequential optimization $W$& Simultaneous optimization $W$\\
    \hline
    $\mathrm{H}_{ss}$ & -0.071550 & -0.020539  & -0.020578 \\
    \hline
    $\mathrm{C}_{ss}$ & -0.030200 & -0.026541  & -0.030650\\
    \hline
    $\mathrm{C}_{sp/ps}$ & -0.025025 & -0.011574  & -0.003557\\
    \hline
    $\mathrm{C}_{pp}$ & -0.022725 & -0.046363  & -0.045573\\
    \hline
    $\mathrm{N}_{ss}$ & -0.033000 & -0.033796  & -0.033939\\
    \hline
    $\mathrm{N}_{sp/ps}$ & -0.027475 & -0.027861  & -0.028144\\
    \hline
    $\mathrm{N}_{pp}$ & -0.025475 & -0.029340  & -0.029484\\
    \hline
    $\mathrm{O}_{ss}$ & -0.035100 & -0.037118  & -0.036986\\
    \hline
    $\mathrm{O}_{sp/ps}$ & -0.029500 & -0.030460  & -0.030598\\
    \hline
    $\mathrm{O}_{pp}$ & -0.027825 & -0.065741  & -0.065757\\
    \hline
    $\mathrm{F}_{ss}$ & -0.036900 & -0.024022  & -0.024519\\
    \hline
    $\mathrm{F}_{sp/ps}$ & -0.031200 & -0.040506  & -0.039640\\
    \hline
    $\mathrm{F}_{pp}$ & -0.029900 & -0.010484  & -0.010632\\
    \hline
  \end{tabular}
\end{table}

\noindent The sequential optimization strategy allowed us to reduce the root mean square error (RMSE) from 49.23 to 15.09 kcal/mol for the W4-11-HCONF subset. In addition, we optimized all the parameters simultaneously, and as expected this optimization resulted in a slightly lower RMSE of 14.88 kcal/mol for the W4-11-HCONF subset. The good agreement of spin-polarization parameters obtained with both approaches signals a minor correlation between the individual parameters as already expected from the agreement between the first- and total-order Sobol indices in Fig \ref{fig:w4_sens}. 

%\textcolor{red}{The sequential approach offers atom-specific parameters, ensuring optimal accuracy when dealing with specific subsets of atoms, while the simultaneous optimization provides a comprehensive set suitable for diverse molecular systems.} \\

\noindent In Table \ref{tab:rmse}, we summarize results for several datasets to analyze the transferability of the W4-11-HCONF-optimized parameters. These datasets include a second set of atomization energies (Alkatom19\cite{karton2009benchmark}), radical-stabilisation energies (RSE43 \cite{goerigk2017look}), hydrogen combustion\cite{guan2022benchmark}, and rearrangements in radical cations (RC21  \cite{goerigk2017look}).

%\noindent \textcolor{blue}{Siyavash: All of the RMSEs for these test datasets are calculated by sequential optimized spin parameters because the simultaneous ones are for when we have all those elements (H, C, N, O, F) in a dataset but here is not the case, for example Alkatom19 only has C and H, so we use the sequential optimized parameters for C and H because those are the parameters that were optimized only for the systems that had C and H.}

\noindent With the sequentially optimized parameters, the RMSE for Alkatom19 was reduced from 30.54 to 14.79 kcal/mol. This substantial improvement points toward a principal transferability of optimized spin-polarization parameters between different datasets. In the case of RSE43, the optimized parameters reduced the error by more than half, from 3.98 to 1.61 kcal/mol. 

\begin{figure}[b!]
        \includegraphics[width=.49\textwidth,height=0.25\columnwidth]{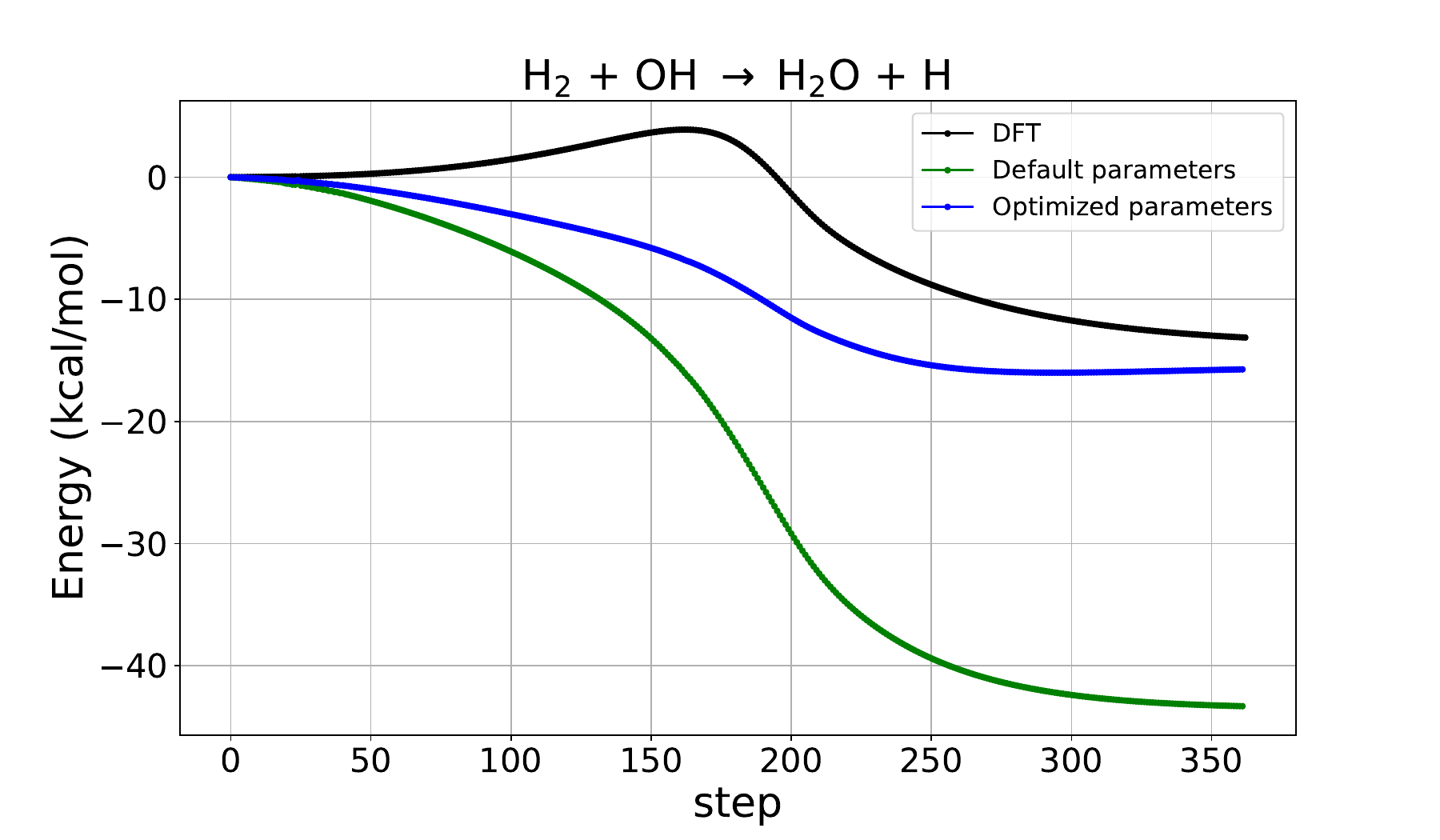}
        \includegraphics[width=.49\textwidth,height=0.25\columnwidth]{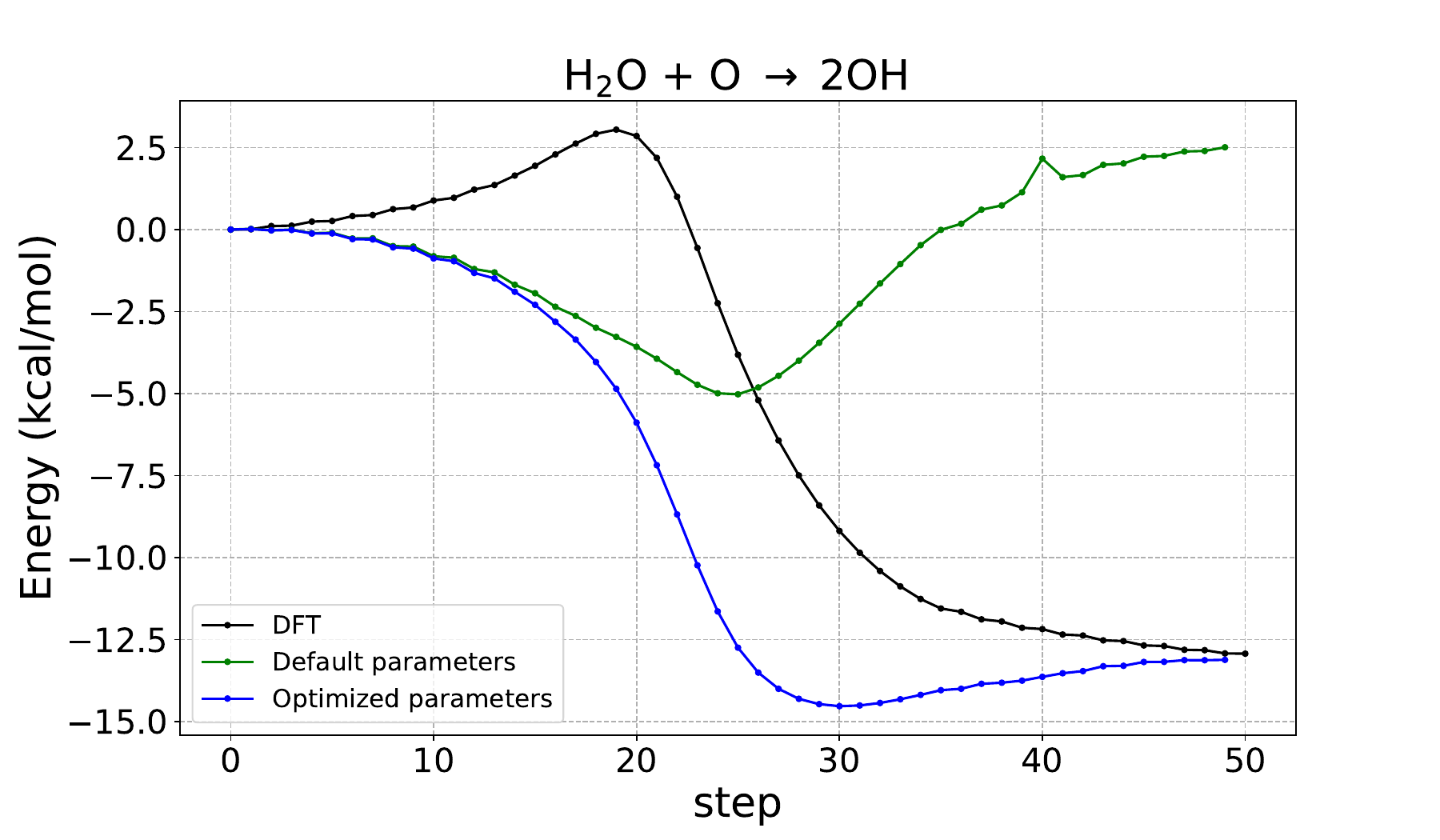}
        \\
        \includegraphics[width=.49\textwidth,height=0.25\columnwidth]{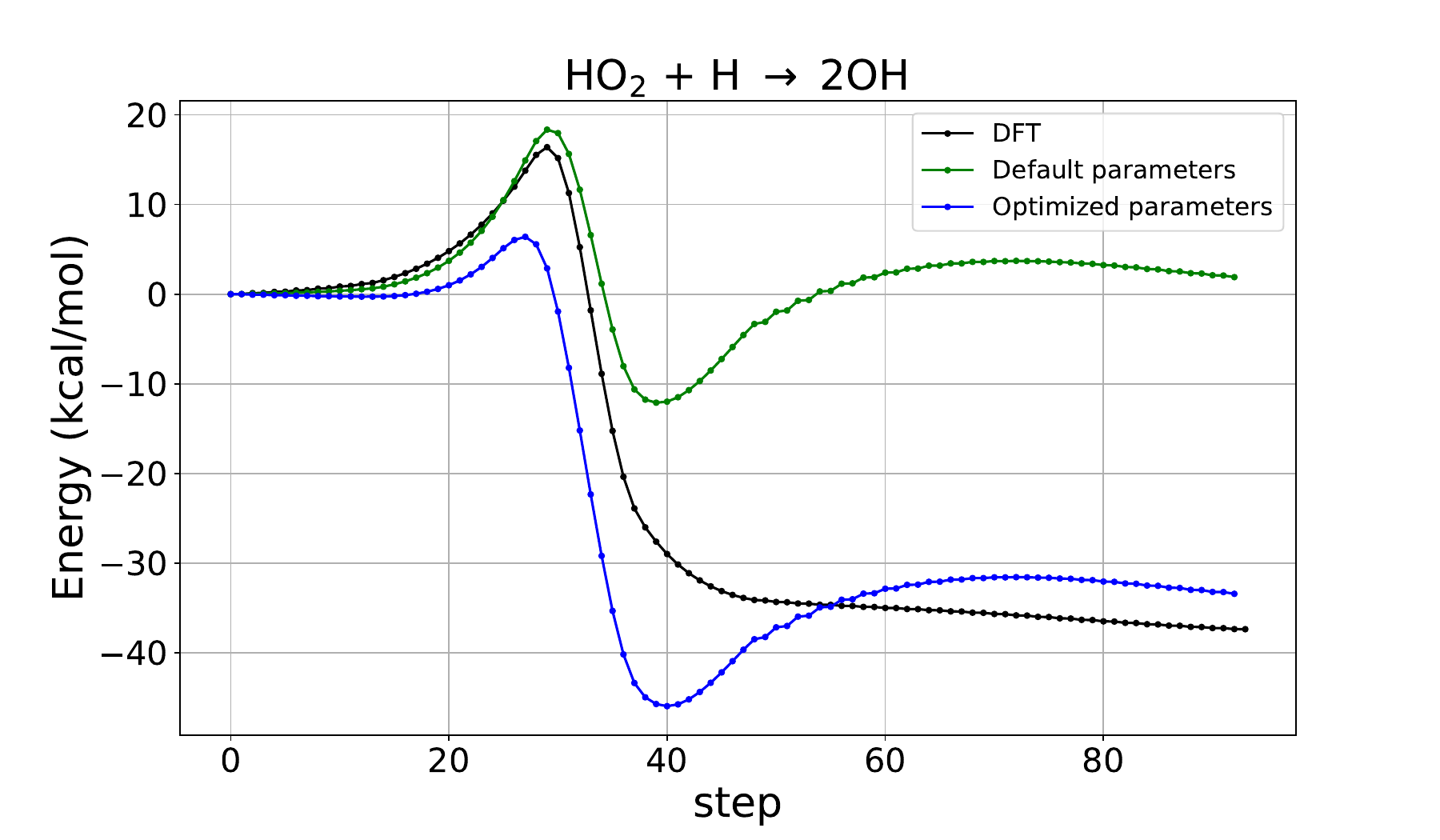}
        \includegraphics[width=.49\textwidth,height=0.25\columnwidth]{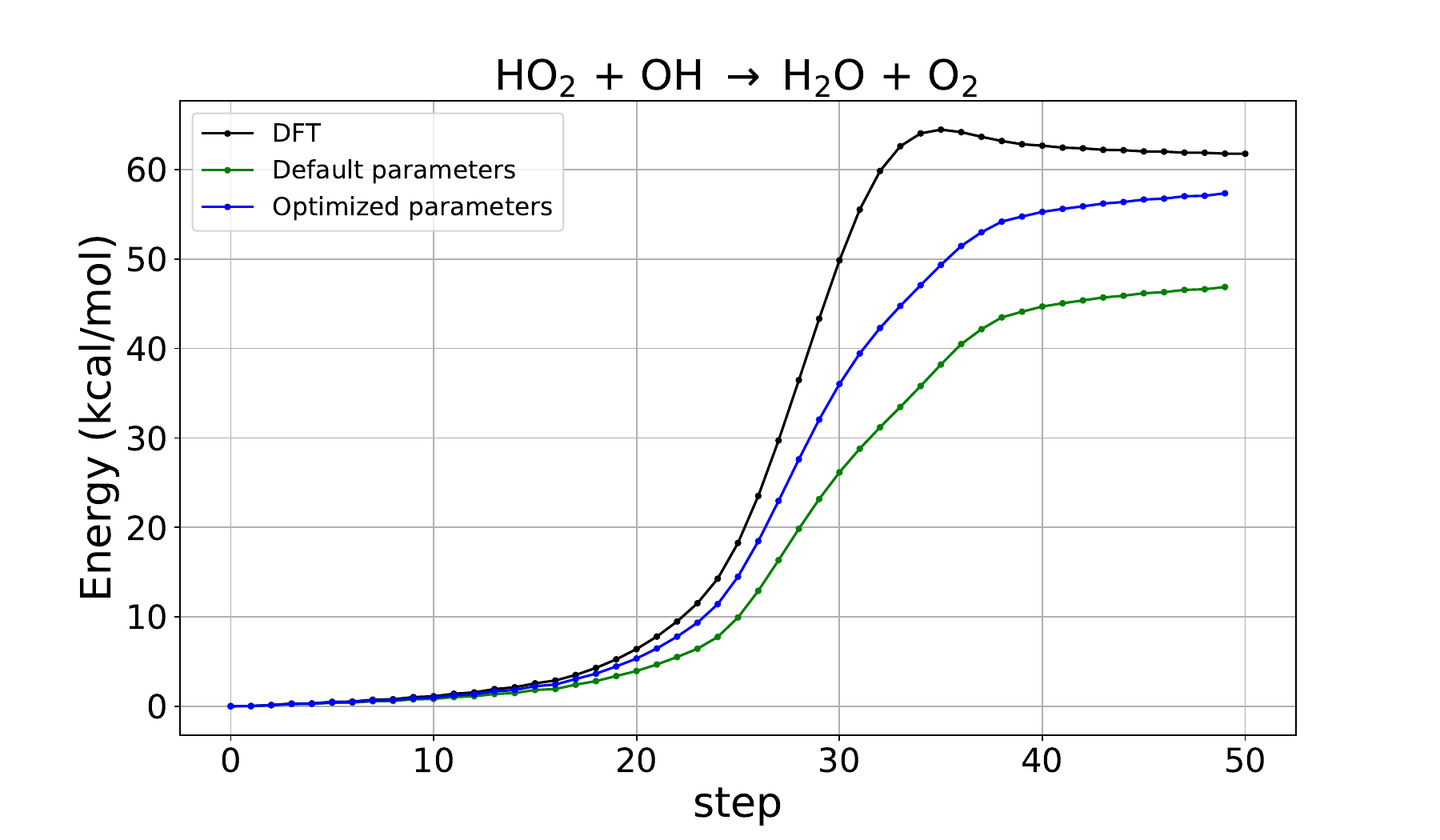}
        \\
        \includegraphics[width=.49\textwidth,height=0.25\columnwidth]{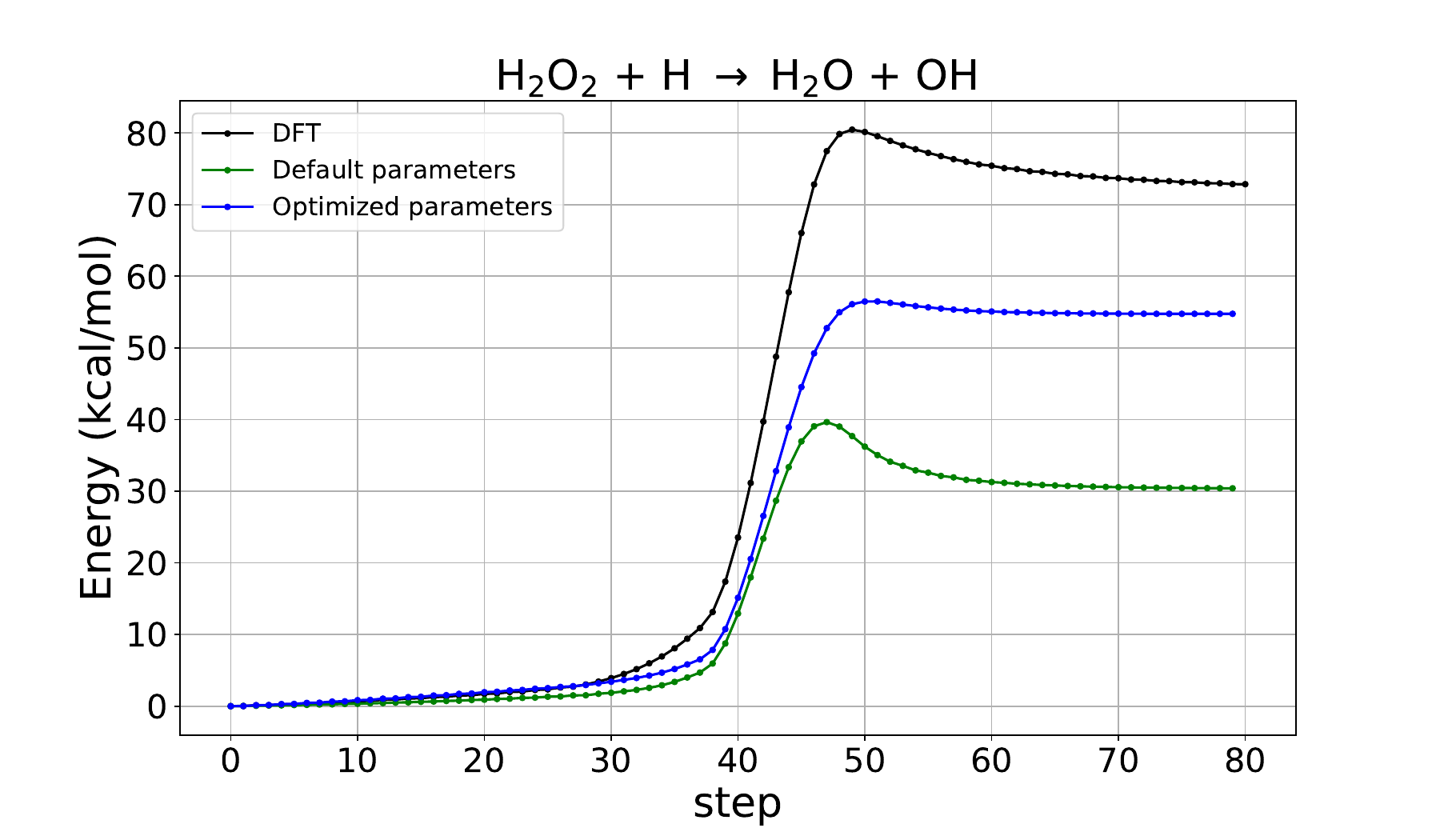}
        \includegraphics[width=.49\textwidth,height=0.25\columnwidth]{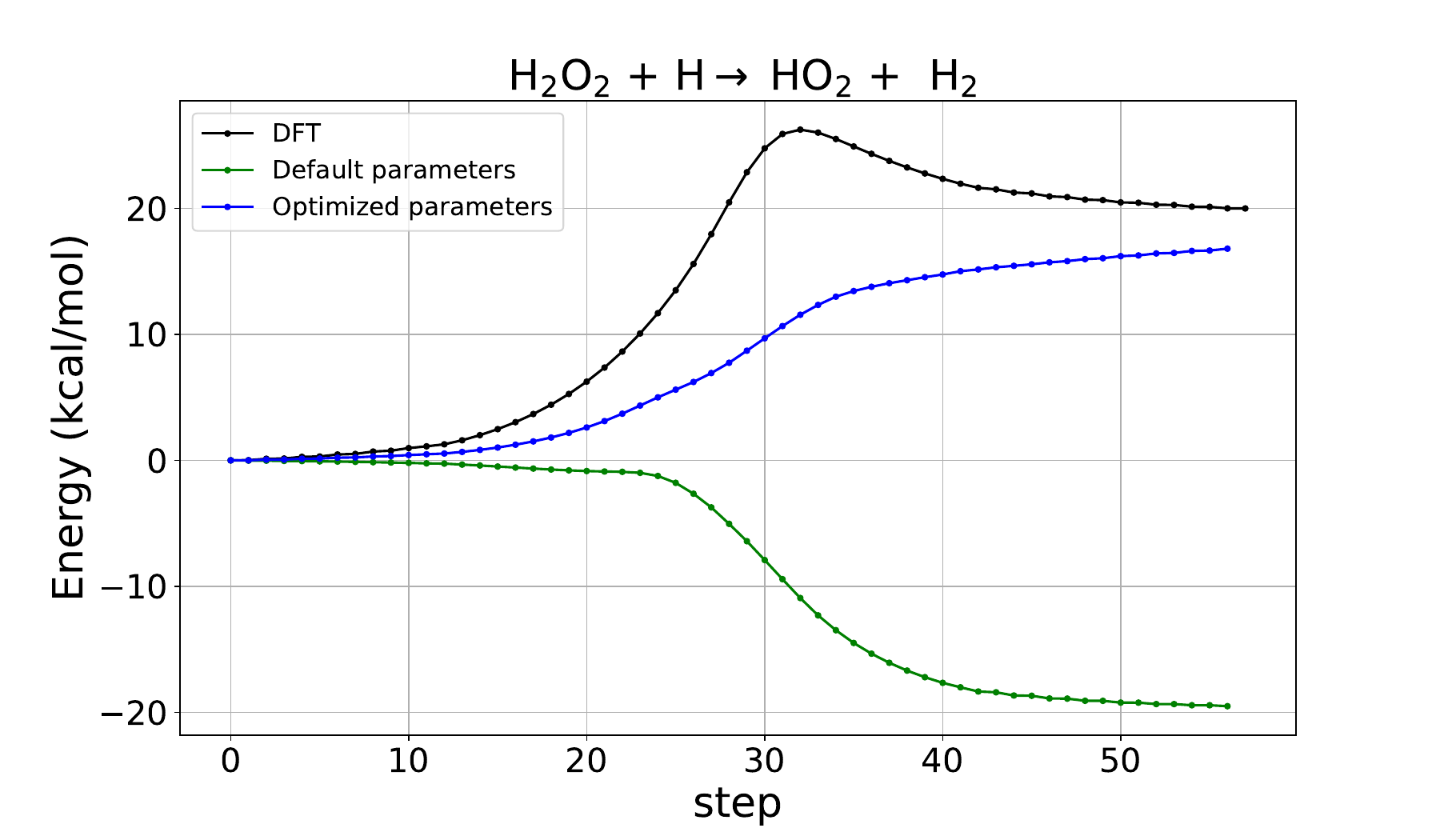}
\caption{\label{fig:irc}
IRCs for several hydrogen combustion reactions with default and optimized spin-polarization parameters. The energy (in kcal/mol) is shown for the DFT reference calculations employing the $\omega$B97X-V functional \cite{mardirossian2014omegab97x} and the cc-pVTZ basis set in black, Q-Chem-xTB with default parameters in green and Q-Chem-xTB with optimized parameters in blue.} 
\end{figure}

\noindent The datasets discussed so far were based on energy differences between minimum structures. An analysis of the performance of spin-polarized GFN2-xTB for non-equilibrium structures is especially useful since this allows us to demonstrate the power of the method for the understanding of chemical reactivity, which the initial method was not parameterized for.
We therefore included a dataset with 19 intrinsic reaction coordinates (IRCs) for hydrogen combustion\cite{guan2022benchmark}. Using the optimized parameters led to an RMSE reduction from 21.96 to 10.50 kcal/mol for the intrinsic reaction coordinates of the hydrogen combustion dataset. 
Comparing the IRC curves obtained by the default and optimized spin parameters in Figure \ref{fig:irc} with the DFT reference data reveals that the generally poor performance of GFN2-xTB can be significantly improved by optimizing the spin-polarization parameters.  
~

However, in the case of radical-cation rearrangements in the RC21 dataset, the error reduction was more modest, from 22.00 to 18.89 kcal/mol. This indicates that while the optimized parameters exhibit transferability across various datasets targeting similar properties, other properties or systems may pose challenges that necessitate further refinement.

\begin{table}
  \centering
  \begin{tabular}{|l|c|c|}
    \hline
      & Default $W$& Optimized $W$\\
    \hline
    W4-11-HCONF& 49.22 & 15.09  \\
    \hline
    Alkatom19 & 30.54 & 14.79  \\
    \hline
    RSE43 & 3.89 & 1.61  \\
    \hline
    Hydrogen Combustion & 21.96 & 10.50  \\
    \hline
    RC21 & 22.00 & 18.98  \\
    \hline
  \end{tabular}
  
  \caption{RMSE in kcal/mol, using spin parameters obtained from sequential optimization of  the W4-11-HCONF dataset}
  \label{tab:rmse}
\end{table}

\noindent 
%CJS: add reference here!
%CIA: fixed

%CJS: You need to call it "Default parameters" rather than "Grimme parameters" in the caption
%CIA: fixed
\subsection{3.2 Limits of transferability} 
\begin{figure}[b!]
    \centering
    \includegraphics[width=\textwidth]{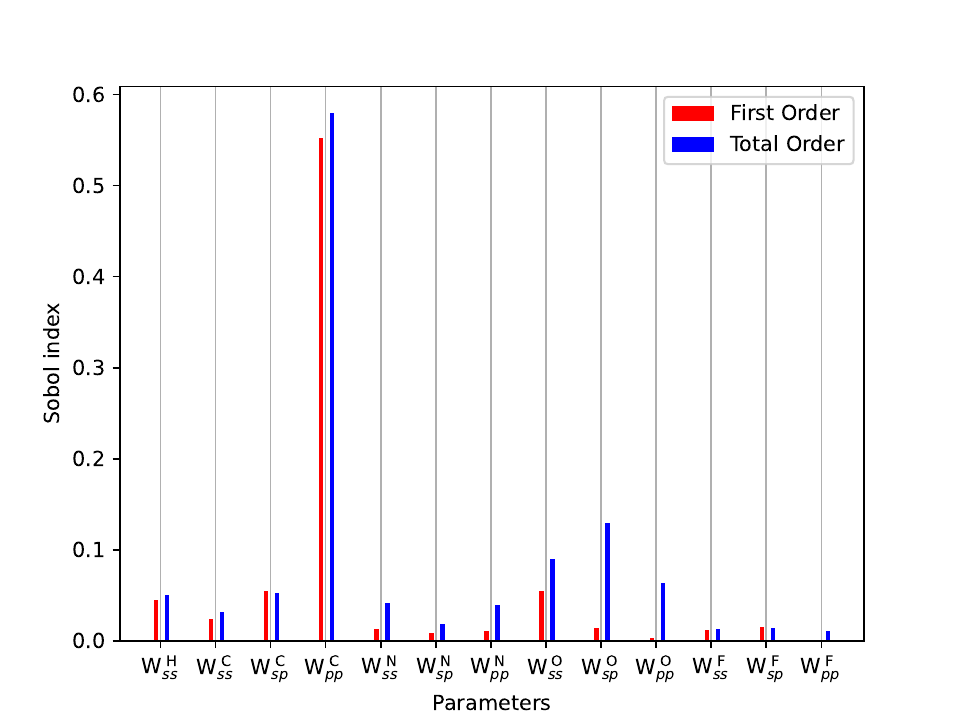}
    \caption{\label{fig:spin_sens} Sensitivity analysis of spin parameters for the AC12 dataset within QChem-xTB.}
\end{figure}

While the transferability of optimized spin-polarization parameters to different datasets has been promising in the cases discussed so far, the W4-11-HCONF-optimized spin-polarization constants yielded poor results when applied to the AC12 dataset of singlet-triplet gaps for a set of aryl carbenes.
\cite{ghafarian2018accurate}. The default parameters yielded an error of approximately 7~kcal/mol, whereas the application of optimized parameters resulted in an unexpected increase of the error to 23.59 kcal/mol. This discrepancy underscores the challenge of achieving transferability across different molecular properties using a single set of optimized parameters in the framework of GFN2-xTB.\\
To understand the cause of this poor performance, we performed a sensitivity analysis specific to the AC12 dataset. As evident from Figure \ref{fig:spin_sens} and in line with our chemical understanding of carbenes, the analysis revealed a key parameter, $W_{pp}$ of carbon, as the sole influential factor in accurately predicting the singlet-triplet gap. 
Addressing this, we optimized only $W_{pp}$ of carbon for this dataset alone. Starting from the previously optimized parameters for W4-11-HCONF $(W_{pp} (\mathrm{C})= -0.046363)$, we reoptimized this single particular to $W_{pp} (\mathrm{C})= -0.015818$, thereby decreasing the error to only 2.96 kcal/mol. This highlights the significant impact of this single parameter on the calculated singlet-triplet gap.\\
%CJS: In my opinion, although we are capable to reduce the error here significantly, I think that the underlying problem here is that we use the same reference occupation for singlet and triplet. This is a general problem within DFTB that we maybe cannot solve here but should briefly discuss.
If, by contrast, we only keep $W_{pp}$ of carbon at the value optimized for the W4-11-HCONF subset and optimize all other parameters, the error, initially at 26.76 kcal/mol, is only insignificantly reduced to 23.19 kcal/mol. 
This analysis underscores the importance of sensitivity analysis in identifying relevant parameters to optimize.\\
The general lack of transferability emphasizes the system-specific nature of the spin parameters within this extended tight-binding method. While it is generally desirable to devise a universal parameter set, our study highlights the potential of tailoring parameters for a given system to yield more accurate results.
~

\noindent The lack of parameter transferability extends to ionization potential (G21IP) and electron affinity (G21EA) datasets, for which we again applied the parameters optimized for the W4-11-HCONF subset.\\
For the G21IP dataset of ionization potentials, the default parameters yielded an error of 201.03 kcal/mol, which decreased to 200.12 kcal/mol for the optimized parameters. Similarly, the G21EA dataset of electron affinities, yielded only a modest error reduction from 96.75 kcal/mol with default parameters to 94.48 kcal/mol with optimized parameters.\\
These catastrophically high errors are obviously not a result of an erroneous description of spin-polarization but rather stem from the minimal basis set used within GFN2-xTB.
The contraction of orbitals upon adding a positive charge to a molecule cannot be captured by a minimal basis set. The same is true for negative charges and the extension of orbitals. These findings are an immediate consequences of the fixed minimal basis set in the GFN2-xTB model, which cannot accurately capture orbital expansions and contractions involved in processes where the charge of a system changes.
While extended tight-binding methods certainly excel in predicting many molecular properties at low computational cost, the accurate prediction of ionization potentials and electron affinities demands larger or more flexible basis sets\cite{muller2023atom}.\\

%CJS: We need to add Grimme's paper on adaptive basis sets here along with others on the same topic
%CIA: added

\section{4. Conclusion}
In conclusion, we showed the significant increase in accuracy that can be achieved by a reoptimization of spin-polarization parameters within GFN2-xTB. The sequential and simultaneous optimization approaches, guided by sensitivity analysis using Sobol indices, demonstrated significant improvements in accuracy, particularly for the W4-11-HCONF subset. However, we encountered challenges concerning the transferability of the optimized parameters for different datasets, especially when different properties were targeted.
The sensitivity analysis proved to be a valuable tool to identify the parameters with the largest impact on the results for a given dataset.
Despite these limits on transferability, we emphasize that a significant error reduction could be achieved by reoptimization for the datasets studied here. This certainly motivates to optimize the spin-polarization parameters based on fits to reference data --- as has been done for almost all parameters within GFN2-xTB --- rather than deriving them from DFT via Janak's theorem.
Importantly, the aforementioned challenges associated with transferability encourage a system- or property-specific optimization rather than aiming for a completely general method. 
For those datasets where transferability was limited, it was found that errors associated with the formulation of the extended tight-binding methods themselves outweighed any potential gains in accuracy associated with parameter optimization; in case of ionization potentials end electron affinities, the lack of flexibility in the the minimal basis set is certainly the biggest source of error, whereas for the AC12 dataset with carbene singlet-triplet gaps, the requirement of DFTB approaches to define a reference occupation to calculate the shell-wise Mulliken charges introduces a bias that leads to an imbalanced description of the two states considered. This is a major limitation of these approaches that motivates further studies that are currently ongoing in our labs. 

\section{Acknowledgements}
We thank Songyuan (Adam) Xu for beginning our xTB implementation during an internship in 2019.

\textbf{Notes}\\
The authors declare the following competing financial interest(s): M.H.-G. is a part-owner of Q-Chem Inc, whose software was used for all developments and calculations reported here.

%\bibliography{mr_1}

\providecommand{\latin}[1]{#1}
\makeatletter
\providecommand{\doi}
  {\begingroup\let\do\@makeother\dospecials
  \catcode`\{=1 \catcode`\}=2 \doi@aux}
\providecommand{\doi@aux}[1]{\endgroup\texttt{#1}}
\makeatother
\providecommand*\mcitethebibliography{\thebibliography}
\csname @ifundefined\endcsname{endmcitethebibliography}  {\let\endmcitethebibliography\endthebibliography}{}

\newpage

\includegraphics[width=\textwidth]{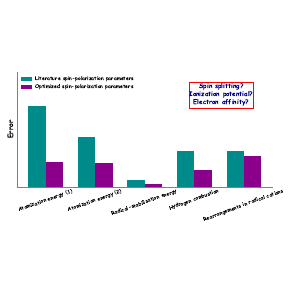}

\end{document}